\documentclass[journal]{IEEEtran}
\ifCLASSINFOpdf
\else
\fi

\usepackage{epsfig,amsfonts,subfigure,amsbsy}
\usepackage{amsthm}
\usepackage[nolist]{acronym}
\usepackage{graphicx,cite,amssymb,amsmath}
\usepackage[usenames,dvipsnames]{color}
\usepackage{tikz}
\usepackage{bigstrut}
\usepackage{psfrag}
\usetikzlibrary{plotmarks}

\usepackage{textcomp}
\usepackage{pgfplots}
\usepackage{mathtools}
\usepackage{etoolbox}
\mathtoolsset{showonlyrefs=true}

\usepackage[abspath]{currfile}[2012/05/06]
\usetikzlibrary{matrix,positioning,shapes,arrows,decorations,calc,fit}
\usetikzlibrary{decorations.pathreplacing}
\usetikzlibrary{spy,backgrounds}
\usetikzlibrary{external}
\usetikzlibrary{patterns}
\usetikzlibrary{shapes,arrows}
\usetikzlibrary{shadows.blur}
\usetikzlibrary{shapes.symbols}
\usetikzlibrary{
	pgfplots.groupplots,
	matrix
}
\allowdisplaybreaks

\usepackage{multirow}
\usepackage{epstopdf}
\usepackage{hyperref}
\graphicspath{{Figures/}}

\IEEEaftertitletext{\vspace{-2\baselineskip}}

\begin{document}
\title{Short Block-length Codes for Ultra-Reliable Low Latency Communications}
\author{Mahyar~Shirvanimoghaddam, Mohamad Sadegh Mohamadi, Rana Abbas, Aleksandar Minja, Chentao Yue, Balazs Matuz, Guojun Han, Zihuai Lin, Yonghui Li, Sarah Johnson, and Branka Vucetic

\thanks{M. Shirvanimoghaddam, R. Abbas, C. Yue, Z. Lin, Y. Li, and B. Vucetic are with the School of Electrical and Information Engineering, The University of Sydney, NSW, Australia.

M. S. Mohammadi is with Silicon Laboratories.

A. Minja is with the Faculty of Engineering (aka Faculty of Technical Sciences), University of Novi Sad, 21000 Serbia.

B. Matuz is with the the Institute of Communications and Navigation of the German Aerospace Center (DLR), Munchner Strasse 20, 82234 Wessling, Germany.

G. Han is with the School of Information Engineering, Guangdong University of Technology, Guangzhou 510006, China.

S. Johnson is with the School of Electrical Engineering and Computing, The University of Newcastle, NSW, Australia.

Corresponding Author: M. Shirvanimoghaddam (email: mahyar.shm@sydney.edu.au)}}
\maketitle

\begin{abstract}
This paper reviews the state of the art channel coding techniques for ultra-reliable low latency communication (URLLC). The stringent requirements of URLLC services, such as ultra-high reliability and low latency, have made it the most challenging feature of the fifth generation (5G) of mobile networks. The problem is even more challenging for the services beyond the 5G promise, such as tele-surgery and factory automation, which require latencies less than 1ms and packet error rates as low as $10^{-9}$. This paper provides an overview on channel coding techniques for URLLC and compares them in terms of performance and complexity. Several important research directions are identified and discussed in more detail.
\end{abstract}
\IEEEpeerreviewmaketitle
\section{Introduction}

\IEEEPARstart{T}{he} third generation partnership project (3GPP) has defined three main service categories in 5G. Enhanced mobile broadband (eMBB) is the service category designed for services which have high requirements for bandwidth such as virtual reality, augmented reality, and high-resolution video streaming.  The second category is massive machine-type communication (mMTC) which  promises to support the massive number of machine-type devices and ultra-low power consumption to increase the device lifetime. 

\begin{figure*}
\centering
\includegraphics[width=2\columnwidth]{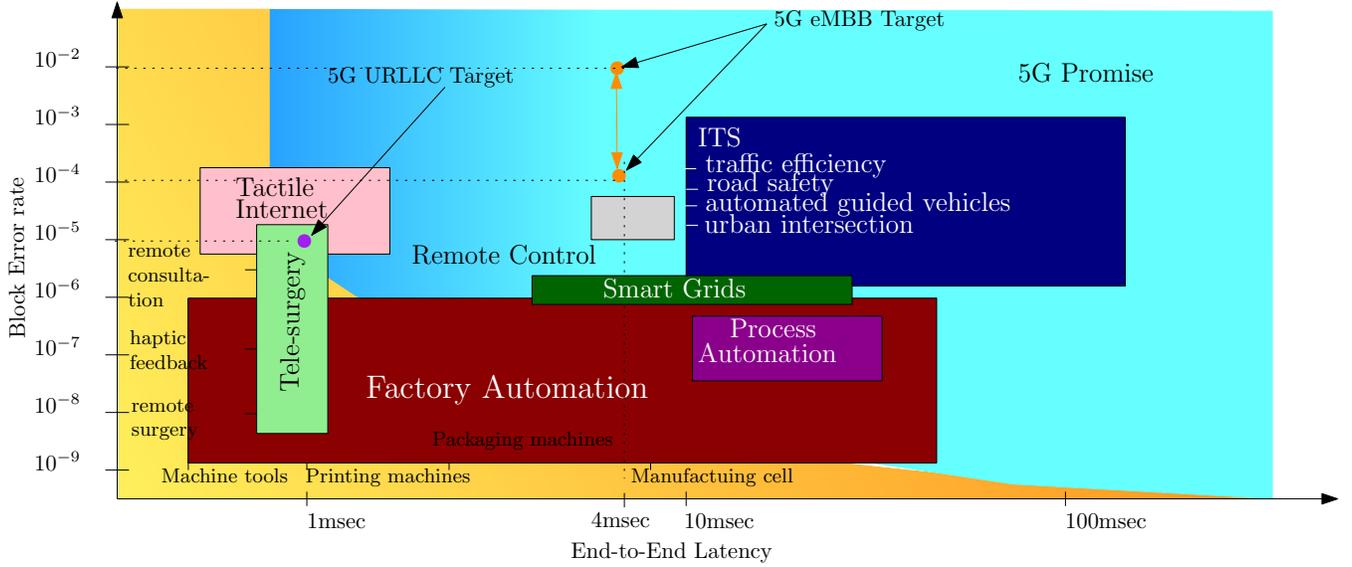}
\caption{Latency and reliability requirements for different URLLC services.}
\vspace{-1.5em}
\label{fig:lr}
\end{figure*}

The third service category is ultra-reliable and low latency communication (URLLC) which focuses on delay sensitive applications and services (see Fig. \ref{fig:lr}). Factory automation and tele-surgery have the strictest reliability requirement of $(1-10^{-9})$ with an end-to-end latency of less than 1ms. Other services such as smart grids, tactile internet, intelligent transportation systems, and process automation have more relaxed reliability requirements of $(1-10^{-3})\sim (1-10^{-6})$ at latencies between 1ms to 100ms  \cite{TS-22.261}. As shown in Fig. \ref{fig:lr}, 5G may not be able to achieve the requirements for some industrial and medical applications with a very strict latency requirement of less than 1ms and block error rate (BLER) of $10^{-9}$. These systems might need to have their own standards with more rigorous latency and reliability levels. For example, power electronics based industrial control needs the overall network latency to be less than 0.1msec and reliability of $(1-10^{-9})$ \cite{Zhibo2018WirelessHP}. 


Physical layer design of URLLC is very challenging because URLLC should satisfy two conflicting requirements: ultra-low latency and ultra-high reliability.  One could use short packets to reduce latency which in turn causes a severe loss in coding gain. Alternatively, the system bandwidth should be widened, which is not always possible especially for some URLLC applications in industrial control that might operate over unlicensed spectrum \cite{Zhibo2018WirelessHP}. On the other hand, for enhancing reliability, we need to use strong channel codes eventually paired with retransmission techniques which indeed increase the latency. 


For high reliability transmissions of URLLC data, a channel code with low code rates is generally used \cite{R1-1804849}. Several candidate channel codes such as low density parity check (LDPC), Polar, tail-biting convolutional code (TB-CC), and Turbo codes, were considered for both eMBB and URLLC data channels. While it was recognized that LDPC , Turbo and Polar codes have similar performance at large block sizes, they could have big performance differences at small block sizes. LDPC has been already selected for eMBB data channel and Polar codes were selected for the eMBB control channel. However, recent investigations showed that there exists some error floors for LDPC codes constructed using base graph (BG) 2, which has been considered for short block low rate scenarios  \cite{R1-1804849}. Moreover, as shown in \cite{R1-1804849}, Polar codes outperform LDPC codes without any sign of error floor. This means that it is not straight-forward to extend/modify eMBB channel coding for URLLC with very diverse latency and data rate requirements. It has been clearly specified that channel coding for URLLC should be further studied, especially for information blocks of less than 1000 bits \cite{R1-1804849}.

In this paper, we compare the main contenders channel codes for URLLC, with the aim to achieve a favorable trade-off between the latency and reliability. We mainly focus on short blocks, i.e., in the order of a few hundreds bits for URLLC. We review existing short channel codes and compare them in terms of rate efficiency at the reliability of interest for some URLLC applications. We show that existing candidate channel codes for URLLC still show a considerable gap to the normal approximation \cite{erseghe2016coding} benchmark, therefore there are still rooms for further improvements.  We highlight several important research directions to improve the performance of URLLC channel codes.

%

\section{Key Metrics, Requirements, and Performance Benchmark}
\subsubsection{Latency}
In the physical layer, we mainly focus on user plane latency, which is defined as the time to successfully deliver a data block from the transmitter to the receiver via the radio interface in both uplink and downlink directions. User plane latency consists of four major components: the time-to-transmit latency, the propagation delay, the processing latency, e.g., for channel estimation and encoding/decoding, and finally the re-transmission time.  Propagation delay is typically defined as the delay of propagation through the transmission medium, and it depends on the distance between the transmitter and receiver. The time-to-transmit latency is required to be in the order of a hundred microseconds, which is much less than the 1ms currently considered in 4G \cite{TS-22.261}. 

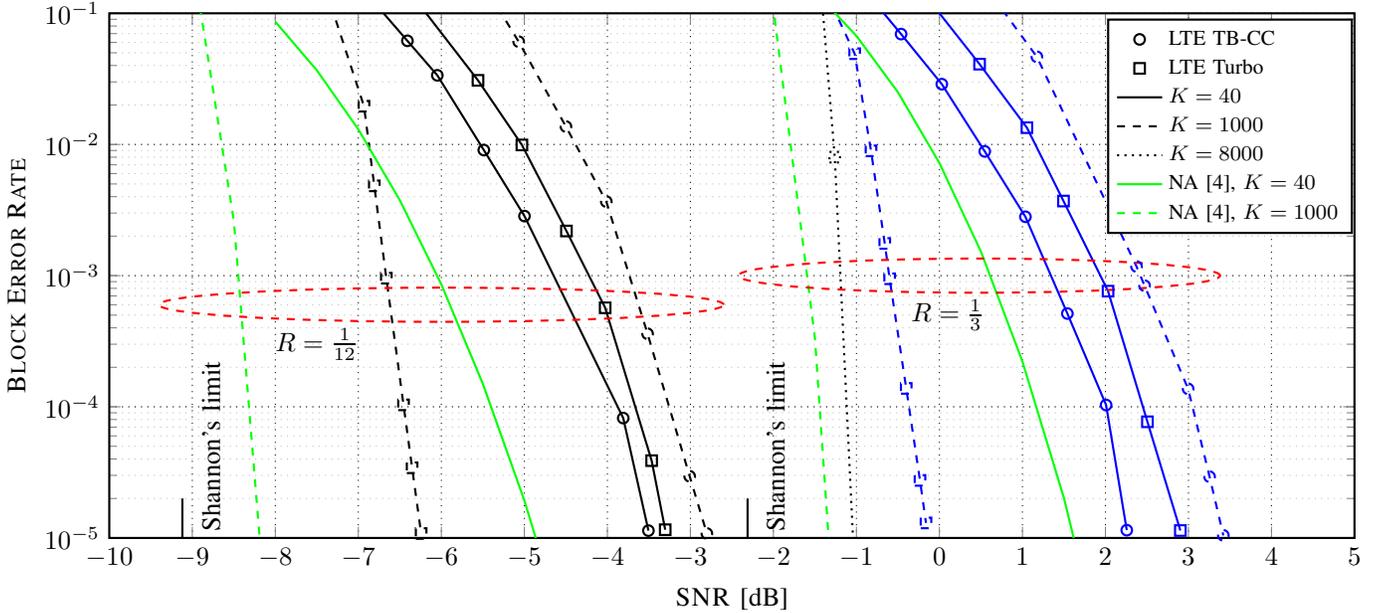
\begin{figure*}
\centering
\begin{tikzpicture}
    \begin{semilogyaxis}[
    thick,
    width=1\textwidth,
height=0.35\textheight,
        xlabel={\textsc{SNR} [dB]},
        ylabel=\textsc{Block Error Rate},
        xmin=-10,xmax=5,
        ymin=1e-5,ymax=0.1,
        grid=both,
        legend style={at={(0.9,0.99)},anchor=north,legend columns=1, legend cell align={left}, font=\footnotesize},
        minor grid style={dotted},
major grid style={dotted,black},
    ]
      \addplot[only marks, mark=o,color=black] plot coordinates {
(-6.950399327448508, 0.15347707446170025)};
      \addplot[only marks, mark=square,color=black] plot coordinates {
(-6.950399327448508, 0.15347707446170025)};

      \addplot[color=black] plot coordinates {
(-6.950399327448508, 0.15347707446170025)};

      \addplot[dashed,color=black] plot coordinates {
(-6.950399327448508, 0.15347707446170025)};

      \addplot[dotted,color=black] plot coordinates {
(-6.950399327448508, 0.15347707446170025)};
%

      \addplot[color=green] plot coordinates {
(-6.950399327448508, 0.15347707446170025)};
      \addplot[dashed, color=green] plot coordinates {
(-6.950399327448508, 0.15347707446170025)};
      \addplot[dotted,color=black] plot coordinates {
(-6.950399327448508, 0.15347707446170025)};

      \addplot[mark=o,color=black] plot coordinates {
(-6.950399327448508, 0.15347707446170025)
(-6.408645088972959  ,0.061689528607400186)
(-6.048339638503574 ,0.03366620067956162)
(-5.488440521227407  ,0.009068962678210345)
(-4.998528793610761  ,0.0028446385080774963)
(-3.808743169398907 , 0.0000820160384886472)
(-3.505464480874317,0.00001137438106253444)
    };
    
   \addplot[mark=o,color=blue] plot coordinates {
   (-1.9113072719630093, 0.42030885872434787)
  (  -1.5224884405212276, 0.29239874686429324)
 (   -0.9874737284573349, 0.17446210838841858)
( -0.46179066834804416, 0.06943030315578953)
(0.028898696931484835, 0.02881918099791986)
(0.5460277427490539, 0.008854919235809087)
(1.035939470365701, 0.002810299931762353)
(1.5414039512400173, 0.0005139502033290993)
(2.0079865489701554, 0.00010316361928018764)
(2.256830601092897, 0.000011466847808767341) 
    
    };
    
\addplot[dashed,mark=o,color=black] plot coordinates {    
 (   -7.894659974147626, 0.9932236714874803)
(-6.859445908666141, 0.8982334573344068)
(-6.714515939498733, 0.833412214377238)
(-5.979513953006877, 0.5304994692069097)
(-5.555076186159468, 0.19412348515788289)
(-5.068525575383169, 0.06078971365556846)
(-4.499157839368352, 0.013691559029900366)
(-4.012607228592053, 0.0036678103620994135)
(-3.5157044771609396, 0.0003604562403399537)
(-3.0084495850750113, 0.000029515052593297577)
(-2.801406771978714, 0.000010808130188758884)

};

\addplot[dashed,mark=o,color=blue]  plot coordinates {
(-1.890418394355006, 0.9932236714874803)
 (-0.059568375689178055, 0.5196572107638463)
(0.5009260969072269, 0.1778040617394324)
(1.1738152394701924, 0.0466100160110624)
(2.374663555428718, 0.0011729011356209648)
(2.4712835348736544, 0.0008381220675038942)
(2.9957919947176084, 0.0001371002499890133)
(3.254595511087981, 0.000029302505957611757)
(3.420229761565018, 0.000010462271211739262)

};

\addplot[mark=square,color=black]  plot coordinates {
(-7.432534678436319, 0.3780396146670867)
(-6.2194199243379575, 0.10715720634368663)
(-5.558427910886928, 0.0308307381927949)
(-5.029634300126104, 0.009917733850621942)
(-4.493064312736444, 0.002189445829414198)
(-4.026481715006305, 0.0005688420390189759)
(-3.466582597730139, 0.000038861758794795056)
(-3.303278688524591, 0.00001156006625297061)
};

\addplot[mark=square,color=blue] plot coordinates {
(-1.9113072719630093, 0.6436539735726591)
(    -0.029424127784782428, 0.1057399411689571)
(0.4838167297183702, 0.04096268533691396)
(1.0540843491663168, 0.01342702278778265)
(1.4947456914670028, 0.00370388623116152)
(2.0313156788566626, 0.0007616463858195501)
(2.505674653215637, 0.0000768720727025861)
(2.902269861286255, 0.000011392814567604962)
};

\addplot[dashed, mark=square,color=black] plot coordinates {
(-7.956772818076516, 0.876027159673047)
(-7.32529223813281, 0.11277098428382175)
(-6.931910893249845, 0.019580484322579604)
(-6.807685205392066, 0.004835773115501188)
(-6.662755236224658, 0.0009562916168201217)
(-6.455712423128361, 0.00010342489876046116)
(-6.352191016580212, 0.000034290485231984576)
(-6.243493539704656, 0.000010808130188758884)
};

\addplot[dashed, mark=square, color=blue] plot coordinates {
(-1.9421790976290811, 0.8640798023397716)
(-1.4349242055431528, 0.17941765490006037)
(-1.020838579350558, 0.049206246852791805)
(-0.8241479069090758, 0.008946561267156372)
(-0.6585136564320369, 0.0017596507284999548)
(-0.5998515260547528, 0.0009465503434081215)
(-0.39971014006166605, 0.00013809471090402836)
(-0.23407588958462888, 0.000027457157583350423)
(-0.16161090500092357, 0.000013018910133482086)
};


\addplot[dotted, mark=square,black] plot coordinates {
(-1.7609,1)
(-1.5109,0.7)
(-1.2609,0.008)
(-1.0109,0.000004)
};

\addplot[ color=green] plot coordinates {
(-8	,0.0862777125938956)
(-7.50000000000000,0.0367745132143350)
(-7	,0.0130400898348836)
(-6.50000000000000,0.00375857763107824)
(-6	,0.000857082692776536)
(-5.50000000000000,	0.000149761070714367)
(-5	,1.93034441355878e-05)
(-4.50000000000000	,1.75375061437305e-06)
(-4,1.06328304527134e-07) 
    };
\addplot[dashed, color=green] plot coordinates {    
    (-10,	0.999850781198246)
(-9.5000000000000,0.931957537274153)
(-9	,0.260345043719065)
(-8.5000000000000,	0.00262893106558353)
(-8,3.44583038367998e-07)
    };
 \addplot[ color=green] plot coordinates {     
(    -5.50000000000000	,0.985038723985394)
(-5	,0.962249935182544)
(-4.50000000000000	,0.917024978505115)
(-4	,0.839627961739964)
(-3.50000000000000	,0.725127334708572)
(-3	,0.578779533959771)
(-2.50000000000000	,0.41787844903052)
(-2	,0.266495660443388)
(-1.50000000000000	,0.146448446274614)
(-1	,0.0673487798279329)
(-0.500000000000000	,0.0250396087228677)
(0	,0.00721358938479177)
(0.500000000000000	,0.00152646438808611)
(1	,0.000221515276176913)
(1.50000000000000	,2.01522471211650e-05)
(2,	1.02024960733376e-06)
};

 \addplot[dashed, color=green] plot coordinates {     
(-3,	0.998501975326310)
(-2.50000000000000,0.813249564006747)
(-2	,0.109596986022233)
(-1.50000000000000	,0.000334514067438268)
(-1,8.21438565792206e-09)
};

\addplot[thick, color=black] plot coordinates {     
(-2.311,1e-5) 
(-2.311,2e-5)
};

\node[anchor=north, rotate=90] at (axis cs: -2.2,0.00005){Shannon's limit};
\addplot[thick, color=black] plot coordinates {     
(-9.12,1e-5) 
(-9.12,2e-5)
};

\node[anchor=north, rotate=90] at (axis cs: -9,0.00005){Shannon's limit};
\draw[dashed, color=red] 
  (axis cs:0.48,0.001) ellipse [ x radius = 29, y radius = 0.3];   

\draw[dashed, color=red] 
  (axis cs:-6,0.0006) ellipse [ x radius = 34, y radius = 0.3];     
  
  \node[] at (axis cs: -7.5,0.0003) {$R=\frac{1}{12}$};
  
  \node[] at (axis cs: 0.1,0.0005) {$R=\frac{1}{3}$};
  
	

    legend style={at={(0,0)}}
    \legend{ LTE TB-CC\\ LTE Turbo\\ $K=40$\\ $K=1000$\\ $K=8000$\\ NA \cite{erseghe2016coding}, $K=40$ \\ NA \cite{erseghe2016coding}, $K=1000$  \\}

    \end{semilogyaxis}
\end{tikzpicture}
\vspace{-1.5em}
    \caption{Comparison of error performance of LTE channel codes with different information block lengths, $K$.}
\label{fig:ppvshort}
\end{figure*}

\subsubsection{Reliability}
Reliability is defined as the success probability of transmitting $K$ information bits within the desired user plane latency at a certain channel quality. Sources of failure from a higher layer perspective are when the packet is lost, or it is received late, or it has residual errors. It is essential to maximize the reliability of every packet in order to minimize the error rate, so as to minimize the number of retransmissions. In this paper, we use block error rate (BLER) as a metric to compare different channel codes in terms of reliability.

\subsubsection{Flexibility}
The flexibility of the channel coding scheme is an important aspect along with the evaluation of the coding performance. Bit-level granularity of  the codeword size and code operating rate is desired for URLLC \cite{R1-1608770}. The actual coding rate used in transmission could not be restricted and optimized for specified ranges \cite{R1-1608770}. The channel codes therefore need to be flexible to enable hybrid automatic repeat request (HARQ). The number of retransmissions however needs to be kept as low as possible to minimize the latency. 

\emph{The general URLLC requirement according to 3GPP is that the reliability of a transmission of one packet of 32 bytes should be $(1-10^{-5})$, within a user plane latency of 1ms (with or without HARQ) \cite{R1-1608770}.} 

\subsubsection{Performance Benchmark}
There are two effects which should be distinguished here to better understand the code design problem for short blocks. The first one is the gap to the Shannon's limit, that is if we decrease the block length, the coding gain will be reduced and the gap to Shannon's limit will increase. This is not a problem of code design but is mainly due to the reduction in channel observations that comes with finite block lengths. We will use the normal approximation (NA) \cite{erseghe2016coding}, that incorporates the reduction in channel observations, as the performance benchmark for comparison. For a coding block of length $N$, the normal approximation is given by \cite{erseghe2016coding}:
\begin{align}
R = C-\sqrt{\frac{V}{N}}Q^{-1}(\epsilon)+\frac{1}{2N}\log_2(N),
\end{align}
where $R$ is the code rate, $C$ is the channel capacity, $V$ is the channel dispersion, $\epsilon$ is the average block error rate (BLER), and $Q(.)$ is the cumulative distribution function of the standard normal distribution. Fig. \ref{fig:ppvshort} shows the normal approximation for different code rates and information block lengths. As can be seen, when the block length increases, the gap to the Shannon's limit \cite{erseghe2016coding} decreases\footnote{It is important to note that Shannon's theoretical model breaks down for short codes, as the channel capacity, defined as the maximum possible rate at which reliable communications is possible, is only valid for infinite block length. The normal approximation was shown to be tight for moderate block lengths ($>$100 bits) \cite{erseghe2016coding}.}.

The second effect is the gap to the finite length bounds, that is if we decrease the block length, modern codes, such as LDPC or Turbo codes, show a gap to finite length bounds. This is often due to the suboptimal decoding algorithms. As can be seen in Fig. \ref{fig:ppvshort}, long term evolution (LTE) Turbo and TB-CC codes show a considerable gap to the normal approximation at short blocks. However, when the block length of the Turbo code increases, the gap to the normal approximation and the Shannon's limit decreases.

\section{Candidate Short Block Length Channel Codes for URLLC}
Here we briefly discuss several fixed-rate channel codes, which might be suitable for URLLC. Throughout the paper, the information block length and codeword length are respectively denoted by $K$ and $N$. For convolutional codes, we use $n$, $k$, and $m$ to denote the bit input and bit output per time instant and memory order, respectively.
\subsection{BCH Codes}
Bose, Chaudhuri, and Hocquenghem (BCH) codes are a class of powerful cyclic error-correcting codes that are constructed using polynomials over finite fields \cite{shuerror}. The main feature of BCH codes is that the number of guaranteed correctable symbols, $t$, is defined during the code design process. The minimum distance $d_{\min}$ of BCH codes is at least $2t+1$ \cite{shuerror}. The decoding of BCH codes is usually done using a bounded distance decoder, like the Berlekamp-Massey algorithm, that can correct any combination of up to $t$ symbol errors.  In order to increase the coding gain, in particular on noisy channels, one may use a soft-input decoder, such as ordered statistics decoder (OSD). 

OSD is a near maximum likelihood (ML) soft decision decoding algorithm for an $(N,K)$ binary linear block code with a given generator matrix. The decoding process consists of three steps. The first step is to reorder the channel output in decreasing reliability order which yields a permutation. The same permutation is applied to the generator matrix. The reordered generator matrix is transformed into a systematic form via Gaussian elimination. The hard decision of the $K$ most reliable values of the channel output are encoded into a codeword via the permuted generator matrix. The reprocessing step consists of generating test error patterns of increasing Hamming weight. This step is repeated until a predefined condition is met and the codeword with the smallest Euclidean distance from the ordered channel output is kept as the best decision. Recent advances in OSD design \cite{wonterghem2} significantly reduce the decoding complexity which makes OSD a good choice for the decoding of short bock length codes.

BCH codes have large minimum distances which avoid flooring the performance at low BLER. However, BCH codes are not flexible as the block length and information length cannot be selected arbitrary. 

\subsection{Convolutional Codes}
Convolutional codes (CC) were first introduced by Elias in 1955 \cite{shuerror}. They differ from block codes as the encoder contains memory. Generally, a rate $R=k/n$ convolutional encoder with memory order $m$ can be realized as a linear sequential circuit with input memory $m$, $k$ inputs, and $n$ outputs, where inputs remain in the encoder for $m$ time units after entering.  Large minimum distances and low error probabilities for convolutional codes are achieved by not only increasing $k$ and $n$, but also by increasing the memory order. The decoding complexity however scales in general exponentially with the memory order in both Viterbi and Bahl, Cocke, Jelinek, and Raviv (BCJR) algorithms \cite{shuerror}. 

When short packets have to be transmitted, terminated convolutional codes represent a promising candidate solution, although the rate loss due to a zero tail termination at short block lengths may be unacceptable. A tail-biting approach \cite{gaudio2017performance} eliminates the rate loss and hence it deserves particular attention when comparing channel codes for short blocks. For these reasons, tail-biting convolutional codes (TB-CCs) are currently considered within the 5G standardization for URLLC. It is worth mentioning that the decoders for TB-CCs are more complex than those for convolutional codes. TB-CC was used in LTE for the broadcast channel and downlink/uplink control information.



\subsection{Turbo Codes}
In 1993, Berrou, Glavieux, and Thitimajshima, introduced Turbo coding, which combines a parallel concatenation of two convolutional encoder and iterative maximum a-posteriori probability (MAP) decoding \cite{shuerror}. Turbo codes have been extensively used for the data channel in LTE. For large blocks, Turbo codes are capable of performing within a few tenths of dB from the Shannon's limit. Unfortunately, Turbo codes with iterative decoding in short and moderate block lengths show a gap of more than 1 dB to the finite-length performance benchmark. LTE Turbo code is known to be well designed for medium block length and code rate $\ge 1/3$. When the code rate and block length are small, LTE-Turbo code performance is degraded.  For Turbo codes, 1-bit granularity is feasible for all coding rates and for full range of block size, and the ability of Turbo codes to support both chase combining  and incremental redundancy HARQ is well known \cite{R1-1611081}. 

\subsection{Low Density Parity Check (LDPC) Codes}
Low-density parity-check (LDPC) codes were originally proposed by Gallager in the early 1960s and later rediscovered in the 1990s, when researchers began to investigate codes-on-graph based on Tanner's work in 1981 and iterative decoding \cite{shuerror}. LDPC codes with iterative belief propagation (BP) decoding have been shown to perform very close to Shannon's limit with only a fraction of a decibel gap. Binary LDPC codes with iterative BP decoding however do not perform well at short to moderate block-lengths which is mainly due to the existence of many short cycles in the code's bipartite graph. Recently, protograph-based LDPC codes have been shown to perform well under belief propagation decoding at short-to-moderate block length, but their performance is not comparable with BCH codes under OSD or TB-CCs with large memory. However, they favor the very low decoding complexity under iterative decoding algorithms. Non-binary LDPC codes are also shown to perform very close to the finite length performance bound where the decoding complexity is the major drawback.  

There are some other advantages for LDPC codes, e.g. in parallelization of the decoding algorithm. LDPC codes have been adopted for eMBB data channel and it is a natural extension to apply LDPC codes for URLLC. Two base graphs are considered for LDPC in eMBB. BG 1 is used for high data rate and long block lengths. BG 2 is used for low code rates and short block lengths. Recent investigations demonstrated that there exist some error floors for LDPC codes at certain rates and block lengths \cite{R1-1611081}. Moreover, the complexity of LDPC increases with increasing flexibility \cite{R1-1611081}.




\subsection{Polar Codes}
Polar codes as introduced in \cite{5075875}, are binary linear codes that can provably achieve the capacity of a binary-input discrete memoryless channel using low-complexity encoding and decoding as the code length tends to infinity. Channel polarization is a central technique in the construction of these codes, in which the block code translates $N$ independent and identical binary-input discrete memoryless channels into $N$ synthesized channels with capacities either (close to) zero or one.  The message is only sent over the set of near-perfect channels, and the unreliable channels are unused; In practice they are assigned constant inputs \emph{a priori} known for both the encoder and decoder (frozen symbols). 

Under successive cancellation (SC) decoding which requires a complexity of $\mathcal O(N\log N)$, and for sufficiently large codeword lengths, the block error probability decays exponentially in the square root of the code length. The recursive nature of the SC decoding may impose a large latency depending on the implementation. 


A major improvement in the decoding performance is achieved by using  successive cancellation list (SCL) decoding which keeps a list of most likely decoding paths at all times, unlike the SC decoder which keeps only one decoding path, i.e., it performs a symbol-wise hard decision at each decoding stage. A significant improvement to SCL is cyclic redundancy check (CRC)-aided SCL (CA-SCL), where the message is encoded with a CRC error detection code, and the result is polar coded \cite{6297420}. In this way, the CRC checksum is used at the decoder side to pick the right decoding path in the list, even if it is not the most probable path.

Polar codes have been selected for short blocks for control channels in eMBB \cite{R1-1611081}. Recent investigations and proposals submitted to 3GPP also demonstrated that Polar codes outperform LDPC codes in short block lengths and low code rates without any sign of error floor; therefore suitable for URLLC use cases  \cite{R1-1611081}.  In Polar codes, 1-bit granularity can be achieved for all coding rates and for full range of block size. However, the implementation complexity of the list decoder increases with increasing list size, especially with larger block sizes \cite{R1-1611081}. 


\begin{figure*}
\centering
\begin{tikzpicture}
    \begin{semilogyaxis}[
    thick,
    width=1\textwidth,
height=0.35\textheight,
        xlabel={\textsc{SNR} [dB]},
        ylabel=\textsc{Block Error Rate},
        xmin=1,xmax=5,
        ymin=1e-7,ymax=1,
        grid=major,
        legend style={at={(0.25,0.37)},anchor=north,legend columns=2,legend cell align={left},font=\footnotesize},
major grid style={dotted,black},
    ]

      \addplot[mark=diamond,color=green] plot coordinates {
(1,0.1449)
(1.5,0.05688)
(2,0.01859)
(2.5,0.006806)
(3,0.002095)
(3.5,0.0007447)
(4,0.0001823)
(4.5,0.00004)
(5,0.000007)
    };
    
       \addplot[dashed, mark=diamond,color=green] plot coordinates {
   (1,0.2)
   (1.5,0.07)
   (2,0.02)
   (2.5,0.003)
   (3,0.00045)
   (3.5,0.000055)
   (4,0.000004)
   (4.5,0.0000003)
    (5,0.00000005)
    };
    
       \addplot[mark=star,color=purple] plot coordinates {
   (1,0.1538)
   (1.5,0.0559)
   (2,0.01538)
   (2.5,0.004031)
   (3,0.0006489)
   (3.5,0.000097)
   (4,0.000015)
   (4.5,0.000001)
    (5,0.00000009)
    };
    
    \addplot[mark=square,color=blue] plot coordinates {    
 ( 1,0.1585)
  (1.5,0.05258)
  (2,0.01874)
  (2.5,0.004909)
  (3,0.001121)
  (3.5,0.0002593)
  (4,0.0000531)
  (4.5,0.00000767)
  (5,0.000001094)
};

\addplot[dashed, mark=square,color=blue]  plot coordinates {
(1,0.1018)
(1.5,0.03459)
(2,0.006782)
(2.5,0.0009822)
(3,0.0001153)
(3.5,0.000008263)
(4,0.00000042)
(4.5,0.00000002)
};

\addplot[mark=triangle,color=blue]  plot coordinates {
(1,0.1064)
(1.5,0.03397)
(2,0.008773)
(2.5,0.001168)
(3,0.0001321)
(3.5,0.00001022)
};


\addplot[mark=o,color=black]  plot coordinates {
(1,0.25)
(1.5,0.06)
(2,0.019)
(2.5,0.004)
(3,0.00085)
(3.5,0.00015)
(4,0.000029)
(4.5,0.000003)
(5,0.0000003)
};

\addplot[dotted, mark=o,color=black]  plot coordinates {
(1,0.12)
(1.5,0.04)
(2,0.01)
(2.5,0.0012)
(3,0.00012)
(3.5,0.000011)
(4,0.000001)
(4.5,0.00000009)
};


\addplot[dashed, mark=o,color=black]  plot coordinates {
(1,0.1)
(1.5,0.035)
(2,0.008)
(2.5,0.001)
(3,0.00009)
(3.5,0.000005)
(4,0.0000002)
(4.5,0.000000007)
};


\addplot[color=red] plot coordinates {
(1,0.12)
(1.5,0.035)
(2,0.009)
(2.5,0.0015)
(3,0.00008)
(3.5,0.000003)
(4,0.00000005)
(4.5,0.000000001)
};

\addplot[dashed, color=black] plot coordinates {
(1,	0.102689677840520)
(1.5,0.0323950953629263)
(2,	0.00689536709410716)
(2.50000000000000	,0.000894735320003527)
(3,	0.0000617391099148112)
(3.50000000000000,	0.00000187521478750642)
(4,	0.0000000191782182808798)
};
 legend style={at={(0,0)}}
    \legend{ Polar Code \cite{channelcodes}\\ Polar Code + CRC-7 \cite{channelcodes}\\ Reed-Muller Code \cite{channelcodes}\\ $\mathbb{F}_{16}$ LDPC Code \cite{channelcodes}\\ $\mathbb{F}_{256}$ LDPC Code \cite{channelcodes}\\ Binary LDPC \cite{channelcodes} \\ TB-CC, $m=8$ \cite{gaudio2017performance}\\ TB-CC, $m=11$ \cite{gaudio2017performance}\\ TB-CC, $m=14$ \cite{gaudio2017performance}\\ eBCH Code \cite{channelcodes}\\ Normal Approximation \cite{erseghe2016coding} \\}

    \end{semilogyaxis}
\end{tikzpicture}
\vspace{-1.5em}
\caption{Comparison of error performance of different rate $R=1/2$ channel codes with codeword length of $N=128$ under MLD \cite{channelcodes}. For TB-CC, we used the circular Viterbi algorithm (CVA).}
\label{fig:WER}
\end{figure*}
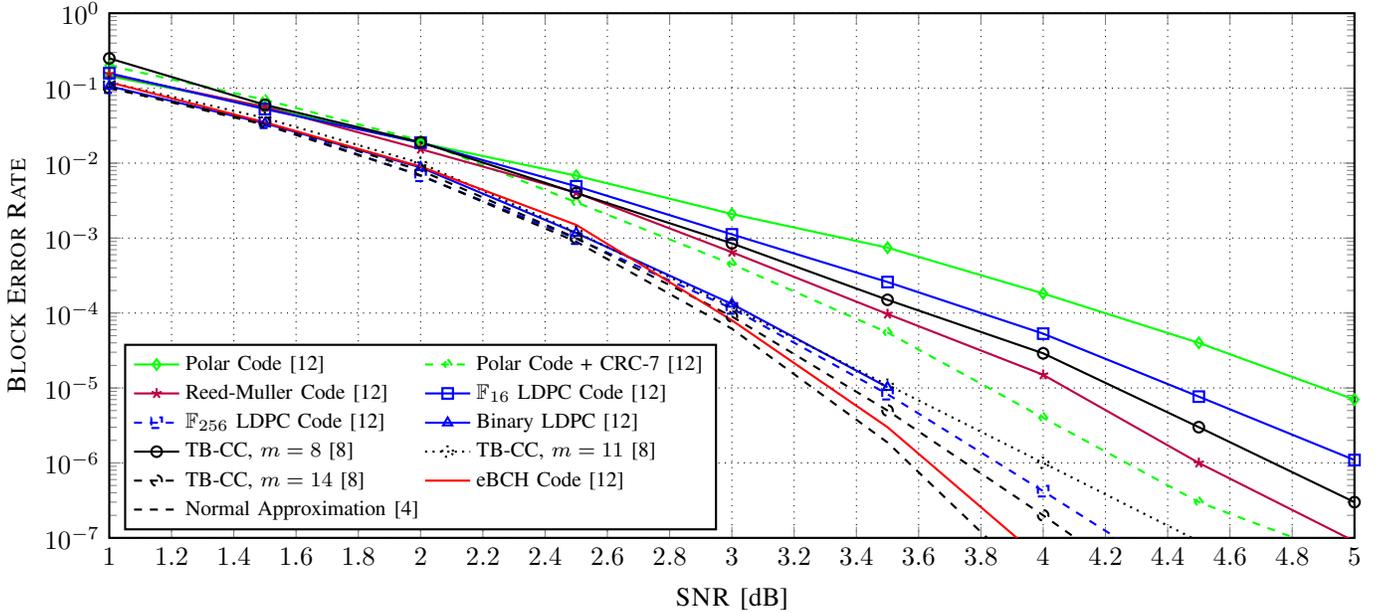
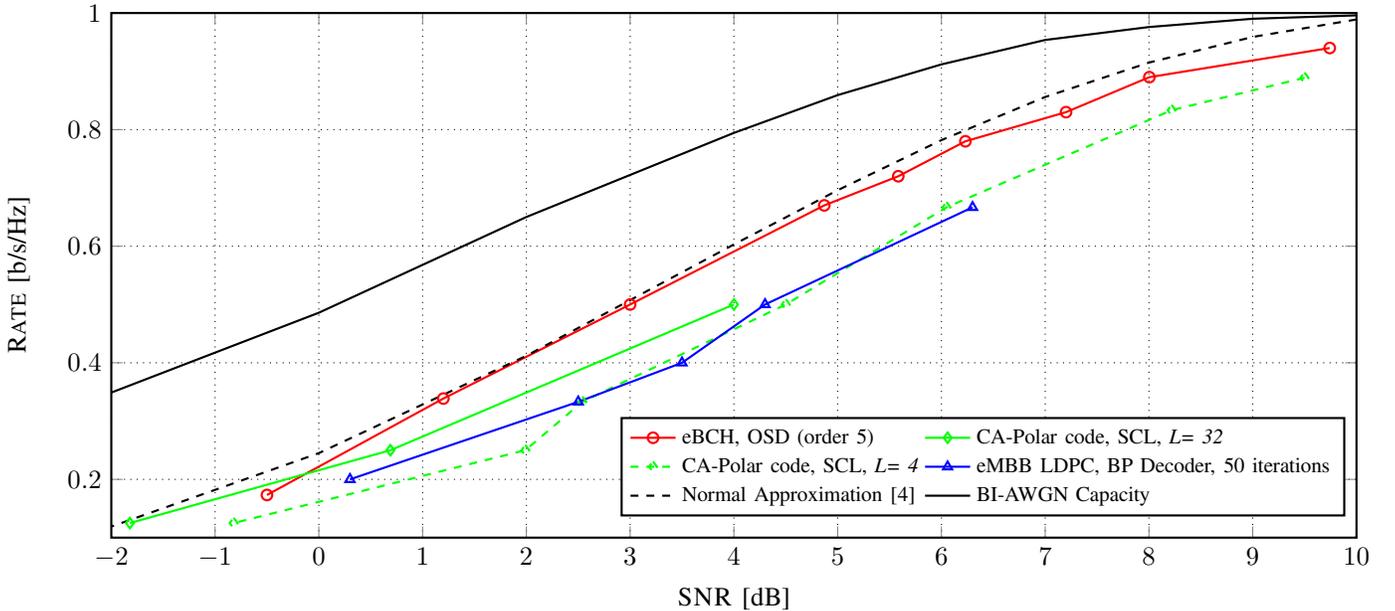
\begin{figure*}
\centering
\begin{tikzpicture}
    \begin{axis}[
    thick,
    width=1\textwidth,
height=0.35\textheight,
        xlabel={\textsc{SNR} [dB]},
        ylabel={\textsc{Rate} [b/s/Hz]},
        xmin=-2,xmax=10,
        ymin=0.1,ymax=1,
        grid=both,
        legend style={at={(0.70,0.23)},anchor=north,legend columns=2,legend cell align={left},font=\footnotesize},
        minor grid style={dotted},
major grid style={dotted,black},
    ]

      \addplot[mark=o,color=red] plot coordinates {
(-0.5,0.1732)
(1.2,0.3386)
(3,0.5)
(4.871,0.67)
(5.584,0.72)
(6.231,0.78)
(7.201,0.83)
(8.004,0.89)
(9.742,0.94)
    };
    
       \addplot[mark=diamond,color=green] plot coordinates {
   (4,0.5)
   (0.6897,0.25)
   (-1.821,0.125)
    };
    
       \addplot[dashed, mark=diamond,color=green] plot coordinates {
   (9.499,0.8889)
   (8.218,0.8333)
   (6.049,0.6667)
   (4.5,0.5)
   (2.539,0.333)
   (1.99,0.25)
   (-0.8206,0.125)
    };

       \addplot[mark=triangle,color=blue] plot coordinates {
   (6.3,0.6667)
   (4.3,0.5)
   (3.5,0.4)
   (2.5,0.333)
   (0.3,0.2)
    };

\addplot[dashed, color=black] plot coordinates {
(10,0.989)
(9,0.959)
(8,0.915)
(7,0.856)
(6,0.782)
(5,0.696)
(4,0.603)
(2,0.412)
(0,0.245)
(-2,0.119)
};

\addplot[thick, color=black] plot coordinates {
(10,0.996)
(9,0.9902)
(8,0.976)
(7,0.9538)
(6,0.9119)
(5,0.8592)
(4,0.7944)
(2,0.6501)
(0,0.4859)
(-2,0.3489)
};

 legend style={at={(0,0)}}
    \legend{ eBCH, OSD (order 5)\\ CA-Polar code, SCL, {\it L= 32} \\ CA-Polar code, SCL, {\it L= 4}\\ eMBB LDPC, BP Decoder, 50 iterations\\  Normal Approximation \cite{erseghe2016coding} \\ BI-AWGN Capacity\\}

    \end{axis}
\end{tikzpicture}
\vspace{-1.5em}
\caption{Comparison of different channel codes with codeword length of $N=128$ with different rates at BLER$=10^{-4}$. }
\label{fig:compare-4}
\end{figure*}
\section{Comparison between Channel Codes for URLLC}
In this part, different channel codes for URLLC are compared in terms of reliability, rate performance, and algorithmic complexity. We consider a binary input additive white Gaussian noise (BI-AWGN) channel, where unit power binary antipodal signals are sent over a channel which are subject to the additive white Gaussian noise of variance $\sigma^2$. The signal to noise ratio (SNR) is then defined as $\frac{1}{\sigma^2}$. For each SNR point and code rate, the simulation is run to obtain 100 codeword errors at the decoder output.

\subsection{Reliability}
Fig. \ref{fig:WER} shows the BLER versus the SNR for different candidate channel codes at rate $R=\frac{1}{2}$ and block length $N=128$ under the maximum likelihood decoding (MLD) \cite{channelcodes}. By using an optimal decoder in the (ML sense), the plot gives insights on the code performance itself. As can be seen in this figure, the extended BCH code closely approaches the normal approximation benchmark over the whole SNR region and can provide a very low BLER as small as $10^{-7}$ with only 0.1dB gap to the normal approximation. Another competitive code is the TB-CC code with $m=14$, which can provide a BLER of  $10^{-5}$ with only 0.1dB gap to the NA benchmark \cite{erseghe2016coding}, however, when it goes to a lower BLER of $10^{-7}$ the gap increases to 0.3dB. Decreasing the memory to 11, TB-CC still gives a performance within 0.1dB gap from the normal approximation at a BLER of $10^{-5}$. Other competitive codes are LDPC codes designed over a large Galois field (here $\mathbb{F}_{\mathrm{256}}$), which have almost the same performance as the TB-CC code with $m=14$. The circular Viterbi algorithm (CVA) algorithm has been used for decoding of TB-CCs \cite{gaudio2017performance}.


The BCH code outperforms all other existing codes owing to its better distance spectrum. Other codes are mainly designed to provide good performance while maintaining the decoding complexity at a reasonable order. We will discuss the tradeoff between complexity and performance in more details in later sections.

\begin{figure*}
\centering
\begin{tikzpicture}
    \begin{semilogyaxis}[
    thick,
    width=1\textwidth,
height=0.35\textheight,
        xlabel={\textsc{SNR} [dB]},
        ylabel=\textsc{Block Error Rate},
        xmin=-10,xmax=5,
        ymin=1e-5,ymax=1,
        grid=both,
        legend style={at={(0.5,-0.15)},anchor=north,legend columns=4,legend cell align={left},font=\footnotesize},
        minor grid style={dotted},
major grid style={dotted,black},
    ]
\addplot[only marks, mark=o,color=black] plot coordinates {           
(  -9.472727272727273, 1.1473647220774152)};

      \addplot[only marks, mark=square,color=black] plot coordinates { 
(-7.5,0.272972973)};

      \addplot[only marks, mark=*,color=black] plot coordinates { 
(-7.5	,0.619631902)};
      \addplot[only marks, mark=diamond,color=black] plot coordinates { 
(-8.4,	0.726618705)};
      \addplot[color=black] plot coordinates { 
(-8.4,	0.726618705)};

      \addplot[dashed,color=black] plot coordinates { 
(-8.4,	0.726618705)};  

      \addplot[dotted,color=black] plot coordinates { 
(-10	,0.999850781198246)
};

      \addplot[mark=o,color=black] plot coordinates {           
(-8	,0.2636)
(-7	,0.0584)
(-6	,0.0052)
(-5	,0.00025)
(-4.6	,5.00E-05)
};
  
      \addplot[mark=o,color=red] plot coordinates {            
(-5	,0.3502)
(-4	,0.1039)
(-3	,0.0133)
(-2	,0.00045)
(-1.6	,7.50E-05)
};
      \addplot[mark=o,color=blue] plot coordinates {
(-2	,0.5512)
(-1	,0.2447)
(0	,0.0548)
(1	,0.0049)
(1.5,	0.00095)
(2,	5.00E-05)
    };
    
      \addplot[dashed, mark=o, color=black] plot coordinates {        
(-9,1)
(-8,	0.2341)
(-7.5,	0.0217)
(-7,	0.0008)
(-6.8, 0.00005)
(-6,0.0000001)
};
      \addplot[dashed, mark=o,color=red] plot coordinates {        
(-5,	0.4558)
(-4.5	,0.0579)
(-4,	0.0021)
(-3.8, 	0.00045)
(-3.5,	0.0000001)
};
      \addplot[dashed, mark=o,color=blue] plot coordinates {        
(-2,	0.9365)
(-1,	0.0545)
(-0.5	,0.0019)
(0,	0.0000500)
};

      \addplot[mark=square,color=black] plot coordinates { 
(-7.5,0.272972973)
(-7,0.131510417)
(-6.5,	0.054742547)
(-6,0.022793952)
(-5.5,	0.006952093)
(-5,0.002049347)
(-4.5	,0.000622465)
(-4,	0.000136266)
(-3.75 ,	0.0000738)
(-3.5,	0.0000294)
(-3,0.00000414)
(-2.75	 , 0.00000152)
};

      \addplot[mark=square,color=red] plot coordinates { 
(-4.75,0.305135952)
(-4.25,0.194980695)
(-3.75,0.088674276)
(-3.25,0.035081626)
(-2.75,0.012609238)
(-2.25,0.004657597)
(-1.75,0.001211481)
(-1.25,0.000286779)
(-0.75,	0.0000704)
(-0.25	,0.0000106)
(0.25,	0.00000177)
};

      \addplot[mark=square,color=blue] plot coordinates { 
(-2.25	,0.497536946)
(-1.75	,0.28611898)
(-1.25	,0.205702648)
(-0.75,	0.092575619)
(-0.25	,0.033926772)
(0.25,	0.017510402)
(0.75	,0.004949767)
(1.25,	0.001284481)
(1.75	,0.000351653)
(2.25,	0.0000678)
(2.75,	0.0000115)
(3.25,	0.00000189)
};

      \addplot[dashed,mark=square,color=black] plot coordinates { 
(-6.2,	0.58045977)
(-5.4	,0.135752688)
(-4.6,	0.017965137)
(-3.8,	0.001795173)
(-3.4,	0.000598097)
(-3,0.000137557)
(-2.6,	0.00000219)
};
      \addplot[dashed,mark=square,color=red] plot coordinates { 
(-2.7,	0.250620347)
(-2.3	,0.110503282)
(-1.5,	0.015277568)
(-1.1,	0.005619228)
(-0.7,	0.001133151)
(-0.3	,0.000328051)
(0.1,0.0000829)
};
      \addplot[dashed,mark=square,color=blue] plot coordinates { 
(0	,0.410569106)
(0.8,	0.107446809)
(1.6	,0.012483006)
(2.4,	0.001057581)
(2.8,	0.000227139)
(3.2,	0.0000546)
};

      \addplot[mark=*,color=black] plot coordinates { 
(-7.5	,0.619631902)
(-6.7	,0.294460641)
(-5.9,	0.121103118)
(-5.1	,0.026288391)
(-4.7,	0.010500052)
(-4.3	,0.00421096)
(-3.5,	0.000641156)
(-3.1	, 0.00021384)
(-2.7,	0.000085)
};
      \addplot[mark=*,color=red] plot coordinates { 
(-5,	0.716312057)
(-4.2,	0.507537688)
(-3.4,	0.242788462)
(-2.6	,0.06746827)
(-1.8,	0.014744526)
(-1.4	,0.005276356)
(-0.6,	0.000674881)
(-0.2	,0.000273207)
(0.2,	0.000111004)
(0.6	,0.000052)
};
      \addplot[mark=*,color=blue] plot coordinates { 
(0	,0.143669986)
(0.4	,0.064045656)
(1.2,	0.015284504)
(1.6,	0.006596134)
(2,	0.001878022)
(2.4,	0.000633328)
(3.2	,0.0001291)
(3.6,	0.0000465)
};

      \addplot[dashed, mark=*,color=black] plot coordinates { 
(-7.1,	0.332236842)
(-6.8	,0.057386364)
(-6.5	,0.001927518)
(-6.2,	0.0000672)
};
      \addplot[dashed, mark=*,color=red] plot coordinates { 
(-4.3,	0.627329193)
(-4,	0.24754902)
(-3.7,	0.022918085)
(-3.4,	0.000744459)
(-3.1,	0.0000404)
};
      \addplot[dashed, mark=*,color=blue] plot coordinates { 
(-1.2,	0.543010753)
(-0.9,	0.121103118)
(-0.6,	0.009477339)
(-0.3,	0.000219623)
(0,	0.0000255)
};

      \addplot[mark=diamond,color=black] plot coordinates { 
(-8.4,	0.726618705)
(-7.6	,0.540106952)
(-6.8	,0.272237197)
(-6	,0.067967699)
(-5.2,	0.020620661)
(-4.4	,0.001661758)
(-3.6,	0.000117102)
(-3.2	,0.0000395)
(-2.4,	0.0000024)
};
      \addplot[mark=diamond,color=red] plot coordinates { 
(-5.2	,0.696551724)
(-4,	0.309815951)
(-3.2	,0.115560641)
(-2.8,	0.060806743)
(-2.4	,0.025325978)
(-1.6	,0.003657035)
(-0.8,	0.000310706)
(0	,0.0000266)
(0.8,	0.00000175)
};
      \addplot[mark=diamond,color=blue] plot coordinates { 
(-2.1	,0.737226277)
(-1.7,	0.554945055)
(-0.9	,0.27520436)
(-0.1,	0.116359447)
(0.7,	0.025811398)
(1.5,	0.003291725)
(2.3	,0.00033768)
(3.1	,0.0000187)
(3.9,0.00000131)
};

      \addplot[dashed,mark=diamond,color=black] plot coordinates { 
(-7.5	,0.5)
(-7.1	,0.058112773)
(-6.9,	0.00844411)
(-6.7,	0.000813177)
(-6.5,	0.0000824)
(-6.1	,0.0000208)
(-5.9,	0.00000826)
(-5.5	,0.00000129)
};

      \addplot[dashed,mark=diamond,color=red] plot coordinates { 
(-4.5,	0.594117647)
(-4.3	,0.215351812)
(-4.1,	0.054831705)
(-3.9,	0.00845825)
(-3.7,	0.000782941)
(-3.5	,0.0000724)
(-3.1,	0.0000119)
(-2.7,0.00000285)
};
      \addplot[dashed,mark=diamond,color=blue] plot coordinates { 
(-1.5,	0.597633136)
(-1.3	,0.279005525)
(-1.1,	0.073241479)
(-0.9,	0.010526316)
(-0.7,	0.001010061)
(-0.5,	0.000110252)
(-0.1	,0.0000130)
(0.5	,0.00000167)
(0.7,	0.000000718)
};
      \addplot[dotted,color=blue] plot coordinates { 
(-4.50000000000000,	0.917024978505115)
(-3.50000000000000,0.72512733470857)
(-2.50000000000000,	0.417787844903052)
(-1.50000000000000,	0.146448446274614)
(-0.500000000000000,	0.0250396087228677)
(0,	0.00721358938479177)
(1,	0.000221515276176913)
(1.50000000000000,	0.0000201)
(2,	0.00000102)
};

      \addplot[dotted,color=red] plot coordinates { 
(-7.50000000000000	,0.823062869636400)
(-6.50000000000000,	0.548551849767591)
(-5,	0.131655035934199)
(-4	,0.0231405968348474)
(-3.50000000000000,	0.00711104282680502)
(-3,	0.00170152490194955)
(-2.50000000000000,	0.000304364749033953)
(-2,	0.0000387199268068245)
(-1.50000000000000,	0.00000329)
};
      \addplot[dotted,color=black] plot coordinates { 
(-8	,0.0862777125938956)
(-7.50000000000000,	0.0367745132143350)
(-7,	0.0130400898348836)
(-6.50000000000000,	0.00375857763107824)
(-6,	0.000857082692776536)
(-5.50000000000000,	0.000149761070714367)
(-5	,0.0000193)
(-4.50000000000000,	0.0000017)
};
      \addplot[dotted,color=blue] plot coordinates { 
(-3	,0.998501975326310)
(-2.50000000000000,	0.813249564006747)
(-2, 0.109596986022233)
(-1.50000000000000,	0.000334514067438268)
(-1,	0.000000008214)
};

      \addplot[dotted,color=red] plot coordinates { 
(-6.50000000000000	,0.995956728698421)
(-6,	0.702579590757071)
(-5.50000000000000,	0.0540635018463238)
(-5,	0.0000790536885023149)
(-4.50000000000000,	0.00000000102525)
};
      \addplot[dotted,color=black] plot coordinates { 
(-10	,0.999850781198246)
(-9.50000000000000	,0.931957537274153)
(-9,	0.260345043719065)
(-8.50000000000000,	0.00262893106558353)
(-8,	0.000000344583038367998)
};

   legend style={at={(0,0)}}
    
    \legend{ CA-Polar, L=32\\ LTE-TB-CC, List-1 Viterbi \\ LDPC, min-sum, 25 iterations \\ LTE-Turbo, Max-log-MAP\\  $K=40$\\$K=1000$\\ NA \cite{erseghe2016coding}\\}

    \end{semilogyaxis}
\end{tikzpicture}
\vspace{-1.5em}
    \caption{Comparison of different channel codes with different rates and information block lengths. Rates $R=1/3$, $1/6$, and  $1/12$, are respectively shown with blue, red and black colors. For the deatil of the channel codes refer to \cite{R1-1611108}.}
\label{fig:urllcshort}
\end{figure*}
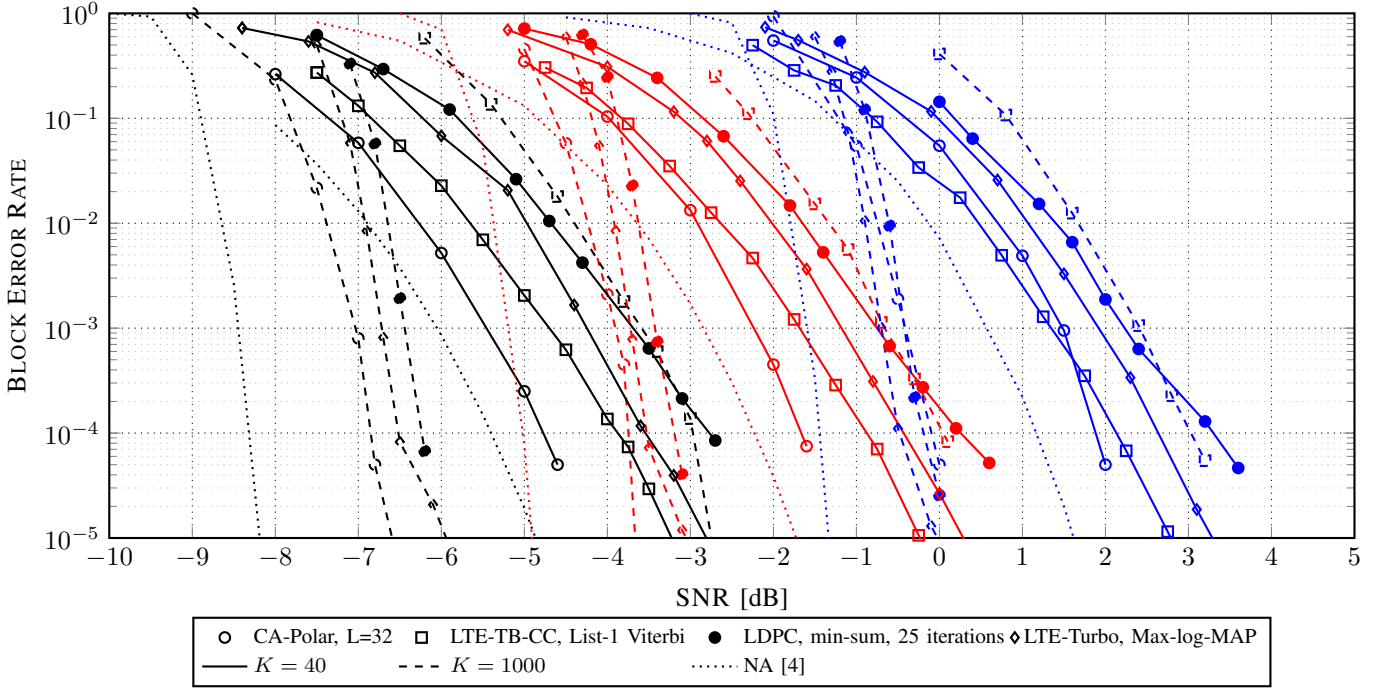
\subsection{Rate Performance}
Fig. \ref{fig:compare-4} shows the rate performance of different candidate codes at a BLER of $10^{-4}$ when the codeword length is $N=128$.  As can be seen, BCH codes perform very close to the normal approximation and outperform other existing codes at all SNRs. The generator polynomials of the used BCH codes were taken from \cite{shuerror} and OSD was used with a maximum re-factoring order of 5. As in \cite{wonterghem2}, we used the probabilistic necessary condition and the probabilistic sufficient condition to reduce the complexity of the OSD decoder.

Two sets of Polar codes with SA-SCL with list sizes 4 and 32 are also shown in Fig. \ref{fig:compare-4}. Although, the decoder with the list size of 32 significantly outperforms the other decoder with list size 4, this comes with a significant increase in the decoding complexity, as the decoder needs to store and list additional candidates.

For the sake of comparison, we also show in Fig. \ref{fig:compare-4} the performance of the short block length LDPC codes specifically designed for eMBB. These codes have the advantage of very low complexity iterative BP decoding and have slightly better performance than the CA-Polar code with SCL decoding of list size 4. 

It is important to note that although BCH codes perform very close to the normal approximation benchmark, their decoding is very complex. For lower code rates, usually an OSD with much higher reprocessing order should be used to guarantee the performance. This however significantly increases the complexity. On the other hand, as can be seen in Fig. \ref{fig:urllcshort}, Polar and LDPC codes can still offer good performance at low rates with considerably lower decoding complexities. 




\subsection{Complexity vs. Performance}
Turbo and LDPC codes have shown to provide near capacity performance at large block-lengths with reasonable complexity due to the iterative nature of the decoders and the fact that most of the calculations can be done in parallel. The complexity of such decoders, for example belief propagation, scale linearly with the block-length. In fact, in most of the complexity analysis of such codes, the complexity is usually characterized in terms of the block length. However, for short block lengths other code parameters have significant impact on the decoding complexity. Here we only focus on the algorithmic complexity which can be represented in terms of the number of binary operations. For example, the decoding complexity of a TB-CC code using the Viterbi decoder is mainly dominated by the memory order, in short block lengths, as the memory order should be usually large to guarantee the performance. 
\begin{figure*}
\centering
\begin{tikzpicture}
    \begin{semilogyaxis}[
    thick,
    width=1\textwidth,
height=0.35\textheight,
        xlabel={\textsc{Gap to the Normal Approximation} [dB]},
        ylabel={\textsc{Number of Binary Operations Per Inf. Bit}},
        xmin=0,xmax=3,
        ymin=10,ymax=1e10,
        legend style={at={(0.5,-0.18)},anchor=north,legend columns=3,legend cell align={left},font=\footnotesize},
        minor grid style={dotted},
major grid style={dotted,black},
    ]

\addplot [only marks,mark=diamond, color=black] coordinates { (2.778,15) };
\node[anchor=west] at (axis cs: 2.4,30){\footnotesize Polar Code, SC};
\addplot [only marks,mark=diamond, color=purple] coordinates { (1.5,68) };
\node[anchor=west] at (axis cs: 1.5,68){\footnotesize Polar Code, SCL, $L=4$};
\addplot [only marks,mark=diamond, color=purple] coordinates { (0.5,640) };
\node[anchor=west] at (axis cs: 0.5,640){\footnotesize Polar Code, SCL, $L=32$};
\addplot [only marks,mark=triangle, color=black] coordinates { (0.8,262144) };
\node[anchor=west] at (axis cs: 0.8,262144){\footnotesize TB-CC $m=8$};
\addplot [only marks,mark=triangle, color=red] coordinates { (0.3333,16777216) };
\node[anchor=west] at (axis cs: 0.3333,16777216){\footnotesize TB-CC $m=11$};
\addplot [only marks,mark=triangle, color=blue] coordinates { (0.1667,1.073e9) };
\node[anchor=west] at (axis cs: 0.1667,1.073e9){\footnotesize TB-CC $m=14$};
\addplot [only marks,mark=square, color=green] coordinates { (0.7778,3.9354e7) };
\node[anchor=west] at (axis cs: 0.7778,3.9354e7){\footnotesize $\mathbb{F}_{256}$ LDPC Code (log-BP)};
\addplot [only marks,mark=square, color=blue] coordinates { (0.7778,5.9354e5) };
\node[anchor=west] at (axis cs: 0.7778,9e5){\footnotesize $\mathbb{F}_{256}$ LDPC Code (FFT-BP)};
\addplot [only marks,mark=star, color=blue] coordinates { (1.25,3550) };
\node[anchor=west] at (axis cs: 0.85,1550){\footnotesize eMBB LDPC BP (50 Iterations)};
\addplot [only marks,mark=o, color=red] coordinates { (1.2,4096) };
\node[anchor=west] at (axis cs: 0.4,5096){\footnotesize Turbo Code, {\it m=4} (Max-log-MAP)};
\addplot[dashed, mark=o,color=black] plot coordinates {
(0.1,1.0737e9)
(0.15,16777216)
(0.23,1.0471e6)
(0.6,5e4)
(1.2,2.5e4)
(1.8,4e2)
(2.75,1e2)
};
\node[anchor=west] at (axis cs: 1.6,3.1e3){\footnotesize eBCH (OSD order 5 with bounded complexity)};
\addplot[thick, color=black] plot coordinates {     
(1.6,1.7e3) 
(1.6,5e3)
};

    \end{semilogyaxis}
\end{tikzpicture}
\vspace{-1.5em}
\caption{Algorithmic complexity versus performance for different rate-$1/2$ channel codes with block length $N=128$ at BLER$=10^{-4}$. The algorithmic complexity for different decoders are obtained from \cite{sybis2016channel}.}
\label{fig_2}
\end{figure*}
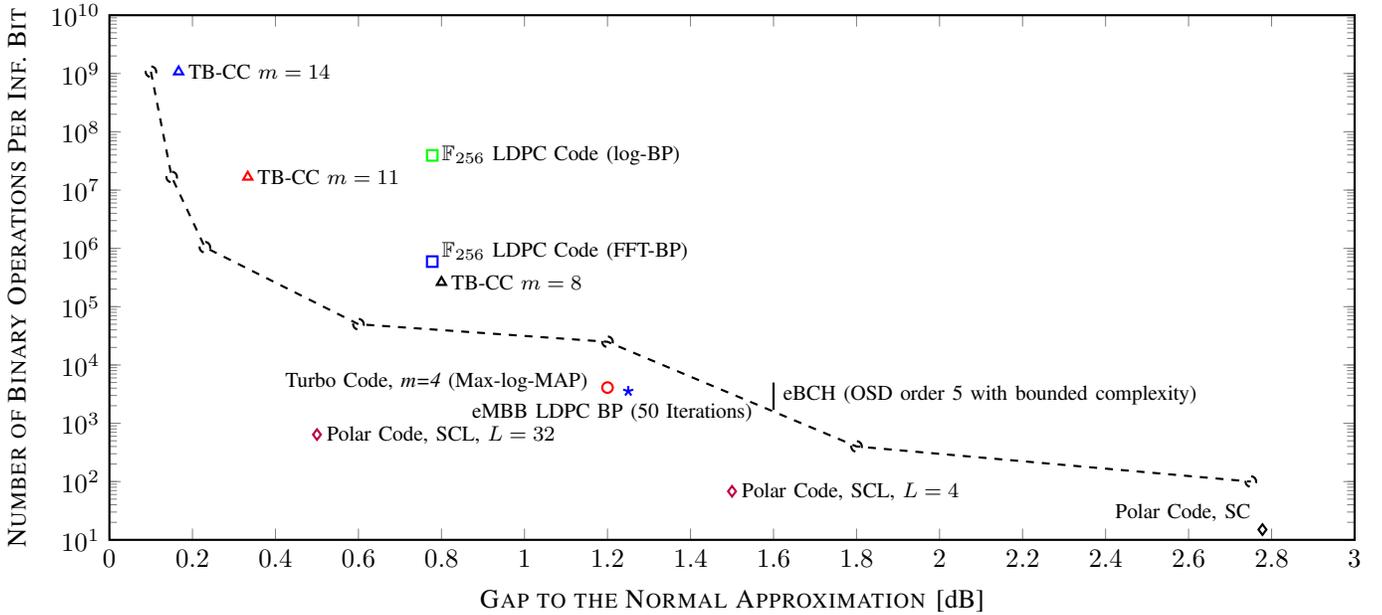
Fig. \ref{fig_2} shows the complexity versus the performance of different channel codes. As can be seen, Polar codes with the SCL decoder achieve the error rate of $10^{-4}$ at only $0.5$dB gap to the normal approximation benchmark with the complexity in the order of $10^3$ operations per bit. The complexity can be reduced by reducing the list size, which however degrades the performance. TB-CC codes have huge complexity which significantly increases with the memory order. The original OSD decoder has the complexity in the order of $K^\ell$, for $\ell$ being the order number. 

\section{Recommendations and Research Directions}
To achieve ultra-high reliability and low latency, we identify three main directions where major improvements are essential. We also provide some recommendations in each direction.

\subsection{Developing Low Complexity OSD Decoders}

As we show in Fig. \ref{fig:compare-4}, under the OSD, BCH codes outperform other existing channel codes, including Polar codes, Turbo codes, and LDPC codes. The complexity of OSD can be significantly reduced using several approaches such as sufficient conditioning and segmentation  \cite{wonterghem2}. Recent work  \cite{wonterghem2} has shown significant complexity reduction for OSD while maintaining rate efficiency similar to the original OSD decoder. Moreover, as shown in Fig. \ref{fig_2} using an OSD with bounded complexity, a tradeoff between performance and complexity can be achieved. We identify that to further improve the performance of channel codes for URLLC more sophisticated ML-like decoders should be designed to allow low complexity decoding of fundamentally better codes in short block lengths, like BCH codes. 




\subsection{Self-Adaptive Joint Coding and Modulation Schemes}
Current cellular networks adopt a rate-adaptive scheme, that is prior to every transmission, the transmitter sends pilot signals to the receiver which enables the latter to estimate the channel state. Then, the receiver feeds back a channel quality indicator (CQI). Based on this CQI, the transmitter selects the best combination of fixed-rate code and modulation scheme from a predetermined set. In case the receiver cannot recover the information (which can be verified with a CRC), it will request a retransmission based on the well-known HARQ protocol.

This rate adaptive scheme suffers from two main drawbacks. The first drawback is due to the choice of channel codes that targets BLERs of $10^{-2}$ making the scheme heavily dependent on retransmissions. In the current LTE standard, each retransmission takes about 7$\sim$8ms, introducing significant delays unacceptable for URLLC. To tackle this, one needs to consider new channel codes more suitable for short blocks and with lower BLERs. Unfortunately, even so, the rate adaptive scheme still suffers from the channel estimation overhead which takes about 5$\sim$8ms in the current LTE standard. This latency can be very costly for mission critical URLLC applications.

Self-adaptive \cite{AFC} channel coding is a promising approach to ensure ultra-low latency transmissions. With self-adaptive codes, the code rate is determined on the fly and automatically adapts to channel conditions, without having any CQI at the transmitter side.  PBRL-LDPC codes offer a fine granularity over information block size and rate. They have been demonstrated to achieve more than 90\% of the NA benchmark and BLERs as low as $10^{-6}$. However, recent investigations show that the performance of PBRL-LDPC codes constructed from a base matrix degrades gradually when more coded symbols are transmitted. For self-adaptive codes, we can identify two main research directions.
\subsubsection{The design of a self-adaptive joint coding and modulation scheme} PBRL LDPC codes are binary codes and the joint design of the code and modulation can significantly improve the performance. As the rateless property of PBRL-LDPC is inspired by Raptor Codes, we propose to replace the Raptor part of the code with rateless codes over real domain, like Analog Fountain Codes (AFC) \cite{AFC}. The joint design of PBRL-LDPC and AFC can offer significant performance improvements.
\subsubsection{Reducing pilot sequence length for channel estimation at the receiver side} For any rateless code, the receiver still needs to know the channel to decode the information. This implies that the transmitter needs to insert pilot symbols into each transmitted block which can incur significant performance losses. These performance losses become more noticeable when the information block size is small. Assuming a powerful (128,64,22) extended BCH (eBCH) code over a BI-AWGN channel with SNR 3.5dB and using only 7 symbols for pilot transmission, our initial results see 5\% loss in the spectral efficiency at BLER $10^{-5}$. The loss increases to 9.8\% and 18\% when a QPSK and 16QAM modulations are considered, respectively. A solution would be to design near ML decoding using OSD without requiring the accurate CSI at the receiver using iterative approaches.

\subsection{Space-Frequency Channel Coding}
%

In current cellular systems, the spatial domain was mainly used to provide multiplexing gain rather than diversity as the main objective was to improve the throughput for moderate reliability order, i.e., BLER of $10^{-2}$ as in LTE. In order to increase reliability, one can use the available transmit and receive antennas to provide spatial diversity rather than spatial multiplexing.  In 4G and 5G,  orthogonal frequency division multiple access (OFDMA) is the major multiple access technology, in which each resource block consists of a number of OFDM symbols. Reducing the number of OFDM symbols per resource block has been identified as an effective approach to reduce the latency. The design of the channel code in these systems is then challenging when considering different diversity sources, including space or frequency. So far, there is no universal framework to design space-frequency channel codes to provide different level of reliability for low latency 

\section{Conclusions}
This paper reviewed the most recent progresses in the design and implementation of short block length channel codes for ultra-reliable low latency communications (URLLC). Several candidate channel codes, including Polar codes, Turbo codes, LDPC codes, Convolutional codes, and BCH codes, were considered and compared in terms of block error rate under optimal decoder, rate performance under practical decoders, and algorithmic complexity of decoding algorithms. BCH codes provide the highest reliability under optimal decoding, since they have the highest minimum Hamming distance. Polar codes with successive cancellation list decoding provide a reliability of $(1-10^{-4})$ with only 0.5dB gap to the normal approximation with reasonable complexity, however better results might be achieved with reduced complexity OSD and BCH codes with the same level of complexity. We also identified several major research directions for channel coding for URLLC.

\bibliographystyle{IEEEtran}
\footnotesize
\bibliography{IEEEabrv,ref}
\end{document}